\documentclass[twocolumn,showpacs,preprintnumbers,amssymb,nofootinbib, aps,prd, eqsecnum]{revtex4-1}
\usepackage{graphicx, booktabs, epsf, epsfig}
\usepackage{bm} 
\usepackage{amsmath}
\usepackage{mathpazo}
\usepackage{color}
\usepackage{ulem}

\definecolor{purple}{rgb}{0.5,0,0.9}

\begin{document}

\title{Searching for black hole echoes from the LIGO-Virgo Catalog GWTC-1}


\author{Nami Uchikata$^1$} 
\email{uchikata@astro.sc.niigata-u.ac.jp}

\author{Hiroyuki Nakano$^2$}
\email{hinakano@law.ryukoku.ac.jp}
\author{Tatsuya Narikawa$^3$}
\email{narikawa@tap.scphys.kyoto-u.ac.jp}

\author{Norichika Sago$^4$}
\email{sago@artsci.kyushu-u.ac.jp}

\author{Hideyuki Tagoshi$^5$}
\email{tagoshi@icrr.u-tokyo.ac.jp}
\author{Takahiro Tanaka$^{3,6}$}
\email{t.tanaka@tap.scphys.kyoto-u.ac.jp}
\affiliation{
$^1$ Graduate School of Science and Technology, Niigata University, Niigata 950-2181, Japan \\
$^2$ Faculty of Law, Ryukoku University, Kyoto 612-8577, Japan \\
$^3$ Department of Physics, Kyoto University, Kyoto 606-8502, Japan \\
$^4$Faculty of Arts and Science, Kyushu University, Fukuoka 819-0395, Japan \\
$^5$Institute for Cosmic Ray Research, The University of Tokyo, Chiba 277-8582, Japan\\
$^6$Center for Gravitational Physics, Yukawa Institute for Theoretical Physics, Kyoto University, Kyoto 606-8502, Japan}

\begin{abstract}%
We have searched for possible gravitational wave echo signals for nine binary black hole merger events observed by Advanced LIGO and Virgo during the first and second observation runs.
To construct an echo template, we consider Kerr spacetime, where the event horizon is replaced by a reflective membrane.
We use a frequency-dependent reflection rate at the angular potential barrier, which is fitted to the numerical data obtained by solving Teukolsky equations.
This reflection rate gives a frequency-dependent transmission rate that is suppressed at lower frequencies in the template.
We also take into account the overall phase shift of the waveform as a parameter, which arises when the wave is reflected at the membrane and potential barrier.
Using this template based on black hole perturbation, we find no significant echo signals in the binary black hole merger events.
\end{abstract}

\date{\today}
\maketitle

\section{Introduction}
Ten binary black hole mergers were observed by Advanced LIGO and Virgo during the first and second observation runs (O1, O2) \cite{gw150914,gw151226,gw170104,gw170814, catalog}.
Waveforms of gravitational waves of binary black hole mergers can be divided into three phases: inspiral, merger, and ringdown.
The ringdown phase is an important part to analyze the properties of the remnant objects.
From black hole quasinormal modes in the ringdown phase, we can estimate the spin and mass of the remnant black holes \cite{echeverria}.
And if we can also observe higher multipole modes, we can test the no-hair theorem of black holes \cite{dreyer}.
So far, the observed ringdown phase signals are not significant enough to test the above issues, although the dominant mode is consistent with the expectation from the inspiral phase within the current detector sensitivity \cite{ligo3}.
There is a proposal to enhance the ringdown data analysis
including overtones of the quasinormal modes \cite{giesler,isi}.

Looking at the data after the ringdown phase, we might be able to tell whether the remnant object is a black hole or a horizonless compact object \cite{cardoso,cardoso2}, such as the gravastar \cite{mazur} or the firewall \cite{almheiri}.
(See Ref.~\cite{cardoso4} for details on testing exotic compact objects.) 
These horizonless objects are considered to be as compact as black holes with a surface located at the Planck scale outside the horizon radius due to quantum modifications.
 Besides the difference in the spacetime structure, 
the most significant difference between black holes and these horizonless objects, compact enough to possess a light ring, in the post ringdown phase is the presence of ''echoes''.
If the event horizon is replaced by a surface, we can expect that the merger-ringdown waveform will be reflected at the surface.
Then the waveform will be partly transmitted at the angular momentum barrier and partly reflected, which will result in observable gravitational wave echoes. 
Abedi et al.~\cite{abedi} have searched for gravitational wave echoes using three binary black hole mergers observed during LIGO O1.
They consider the simplest model, the horizon is replaced by a reflecting membrane at $\sim$ Planck proper length outside the event horizon radius in Kerr spacetime.
They reported echo signals at 3$\sigma$ significance (0.011 in p-value) from 32-second data around each binary black hole event.
However, Asthon et al.~\cite{ashton} have pointed out some problems in the analysis done by Abedi et al.~and Westerweck et al.~\cite{westerweck} have improved the background estimation using 4096-second data around each binary black hole event, which gave lower significance, 0.032 in p-value, than in Abedi et al..
Using the same template waveform given in Abedi et al.~, Nielsen et al.~\cite{nielsen} and Lo et al.~\cite{lo} have also shown lower significance on echo signals evaluated by the Bayes factor using Bayesian analysis, where Lo et al.~have included the inspiral-merger-ringdown waveform into the template as well.
Injection studies given in \cite{westerweck, nielsen, lo} have shown that echo signals with amplitudes of the first echo larger than about 15\% of the merger amplitude are detectable within the current detector sensitivity.
Also, a morphology-independent search shows a lower significance on echo signals \cite{tsang,tsang2}.

Improvement of the evaluation of the significance is important, but we can also improve the template waveform used in Abedi et al.
If the spacetime outside the reflecting surface is entirely Kerr spacetime, we can exactly calculate the reflection rate and the phase shift due to the reflection at the potential barrier, while the reflection rate is assumed to be a frequency-independent parameter and the overall phase shift is fixed to $\pi$ in Abedi et al.
The frequency-dependent reflection rate and the phase shift at the potential barrier were calculated numerically by Nakano et al.~\cite{nakano} in this setup.
The reflection rate also gives a frequency-dependent transmission rate, which affects the template waveform as well.
In this study, we analyze the echo signals using this reflection rate, which is fitted for $0.6 \le q \le 0.8$, where $q$ is the nondimensional Kerr parameter.
Although the phase shift at the barrier can be calculated exactly, that at the reflective membrane is uncertain, i.e., model dependent.
Here we leave the frequency-independent overall phase shift as a free parameter. 
In summary, the construction of a template waveform is the same as Abedi et al., aside from the treatment of the reflection rate and the phase shift as mentioned above.
Based on this template, we search for gravitational wave echo signals for  binary black hole merger events observed by LIGO and Virgo during O1 and O2.
And we use 4096-second data for each event to perform background estimation, adopting the same method done in Westerweck et al.

In this study and in the previous studies mentioned above, a perfect reflection at the membrane is assumed.
However, the reflection at the membrane is also model dependent.
Template waveforms for a more general reflection rate at the membrane are considered in \cite{mark,testa,maggio}, 
and phenomenological templates are proposed in \cite{maselli}.
The validity of the constant echo interval is discussed in \cite{ytwang,ytwang2}.
Recent studies also provide models of gravitational wave echoes based on black hole area quantization \cite{cardoso3}, and quantum black holes \cite{wang,oshita}.

The paper is organized as follows.
In Sec.~II, we explain the template waveform used in our analysis.
In Sec.~III, we describe the method of analysis to evaluate the significance of echo signals using open LIGO data.
In Sec.~IV, we show the results of our analysis. 
We use p-values to evaluate the significance.
Conclusions and discussions are given in Sec.~V. 
\section{Template waveform based on black hole perturbations}
In this study, we consider  a situation in which the spacetime is entirely Kerr spacetime but a reflective membrane is located at about Planck length away from the event horizon radius.
In such a case, after binary black holes merge, the merger-ringdown part of the waves will be  partly reflected both at the membrane and at the angular potential barrier of the Kerr spacetime.
When the waves are reflected at the potential barrier, part of them will be transmitted through the barrier to escape to infinity, which we may observe as gravitational wave echoes.
Therefore, echo waveforms are basically characterized by the reflection rates at the membrane and the potential barrier, and the time interval of echoes $\Delta t_{\rm echo}$, which corresponds to twice the proper distance between the membrane and the potential barrier.
We assume a perfect reflection at the membrane, which is the same assumption as given in \cite{abedi}; that is, we take into account only  the reflection rate at the potential barrier. 

Abedi et al.~\cite{abedi} assumed that the reflection rate at the potential barrier is given by a frequency-independent parameter.
However, if we assume the spacetime is entirely Kerr spacetime, the reflection rate can be calculated by solving the perturbation equations with appropriate boundary conditions, i.e., only outgoing waves at spatial infinity and total reflection of waves at the membrane \cite{nakano}.
The reflection rate obtained from black hole perturbations depends on frequency.
In this paper, we use the reflection rate $R(f)$ given in \cite{nakano},
\begin{widetext}
\begin{equation}
R(f) \approx
\begin{cases}
\displaystyle
     \frac{1+e^{-300(x+0.27-q) } +  e^{-28(x- 0.125- 0.6q) } }{ 1+e^{-300(x+0.27-q) } +  e^{-28(x- 0.125- 0.6q) } + e^{19(x- 0.3- 0.35q) } } & (f >0), \\ \\
 \displaystyle
    \frac{1+e^{-200(|x|- 0.22+ 0.1 q) } +  e^{-28( |x |- 0.39+0.1q) } }{ 1+e^{-200(|x|- 0.22+ 0.1 q) } +  e^{-28( |x |- 0.39+0.1q) } +e^{16( |x |- 0.383+0.09q) }  }    & (f<0).
\end{cases}
\label{Rf}
\end{equation}
\end{widetext}
\noindent 
Here $x=2 \, \pi M f$ and $q=a/M$ with black hole spin $a$ and mass $M$ in $c=G=1$ units.
Reflection rate $R(f)$ in Eq.~\eqref{Rf} is a fit for the numerically calculated one for $0.6 \le q \le 0.8$, in which the remnant spin of the binary black holes observed by LIGO and Virgo varies.
The time interval between neighboring echoes $\Delta t_{\rm echo}$ is evaluated following the formalism given in \cite{abedi}
\begin{equation}
\begin{split}
\Delta t_{\rm echo} = 2 \int_{r_++\Delta r} ^{r_{\rm max}} \frac{r^2+a^2}{r^2-2M r + a^2} dr \, ,
\end{split}
\end{equation}
where $r_{\rm max}$ is the peak of the angular momentum barrier and $\Delta r$ is the location of the membrane away from the horizon, $r_+$.
In the previous studies \cite{abedi,westerweck,nielsen,lo}, the frequency-independent reflection rate and $\Delta t_{\rm echo}$ are assumed to be parameters.
In our case, since both the reflection rate and $\Delta t_{\rm echo}$ depend on $a$ and $M$, we set $(a, M)$ as parameters instead of $[R(f),\Delta t_{\rm echo}]$.
Then the echo template waveform, including $N$ echoes, in the frequency domain $\tilde{h} (f)$  is given by
\begin{equation}
\begin{split}
\tilde{h} (f) = & \sqrt{1-R^2(f)}\, \tilde{h}_0(f) \\
& \times \sum^{N}_{n=1} R(f)^{n-1} e^{- i \left [2\pi f \Delta t_{\rm echo}+ \phi(f) \right ] (n-1)}  \, , 
\end{split}
\label{template1}
\end{equation}
where $\phi(f)$ is the overall phase shift due to the reflections at the membrane and the potential barrier and $ \tilde{h}_0(f)$ is  the Fourier transform of a time-domain waveform, $h_0(t)$, defined below.
Note that $\sqrt{1-R^2(f)}$ is the transmission rate at the barrier.
We choose $h_0(t)$ as
\begin{equation}
\begin{split}
 h_0(t) &= \frac{1}{2}\left \{ 1 + \tanh \left [ \frac{1}{2}\omega(t) (t-t_{\rm merger} -t_0) \right ] \right \} \\
     & \quad  \times h_{ \mbox{{\tiny IMR}} }(t) \\
 &  \equiv  \Theta(t;t_0,\omega)\, h_{ \mbox{{\tiny IMR}} }(t),
 \end{split}
 \label{seed}
\end{equation}
where $h_{ \mbox{{\tiny IMR}} }(t)$ is the inspiral-merger-ringdown waveform for each event, $\Theta(t;t_0,\omega)$ is a cutoff function given in \cite{abedi}, and $t_{\rm merger}$  is the merger time of the binary black hole.
The choice of $h_{ \mbox{{\tiny IMR}} }(t)$ is shown in Sec.~III A. 
The cutoff function is determined by a cutoff parameter $t_0$ and the typical frequency around the merger time $\omega$.
The cutoff parameter is also given in \cite{abedi,westerweck,nielsen,lo}; however, since it is insensitive to the signal-to-noise ratio (SNR) defined in the next section, to save computational cost, we set $t_0$ as a constant. 
Following the best fit value of $t_0$ obtained in \cite{abedi}, we set $t_0=-0.084 \, \Delta t_{\rm{echo}}$ for GW150914 and $t_0=-0.1 \, \Delta t_{\rm{echo}}$ for the rest of the events.
The phase shift at the potential barrier can also be calculated from Teukolsky equations \cite{nakano}.
However, since the phase shift at the membrane is highly model dependent, we take the overall phase shift $\phi(f)$ as a parameter as well.
Basically, $\phi(f)$ depends on the frequency; however, we can approximate it as a linear function $\phi(f)=\phi_0+\phi_1 f$ \cite{nakano}.
Then, the coefficient of the linear part $\phi_1$ can be absorbed by the parameter $\Delta t_{\rm echo}$, and we only need to consider the zeroth order coefficient $\phi_0$.
For $q=0.7$, the frequency dependence of $\phi(f)$ due to the reflection at the barrier is shown in Fig.~1 in Ref.~\cite{nakano}, and  we can see that $\phi(f)$ only weakly depends on frequency around the quasinormal mode frequency.
This fact will partly justify replacing $\phi(f)$ in Eq.~\eqref{template1} with a constant parameter $\phi_0$.
We stress that one important difference from previous studies is that
$h_0(t)$ here must be the complex template having the unobserved
polarization mode in the imaginary part. If we restrict the phase
shift $\phi$ to $0$ or $\pi$, the other polarization mode does not affect
the echo signal, and hence the imaginary part of $h_0(t)$ is unnecessary.

In summary, three parameters $(a,M,\phi_0)$ are considered in the template in our analysis. 
The template is 32 seconds long, including 20 echoes for GW170729 and 30 echoes for the other events.

\begin{figure}[h]
\includegraphics[scale=1.1]{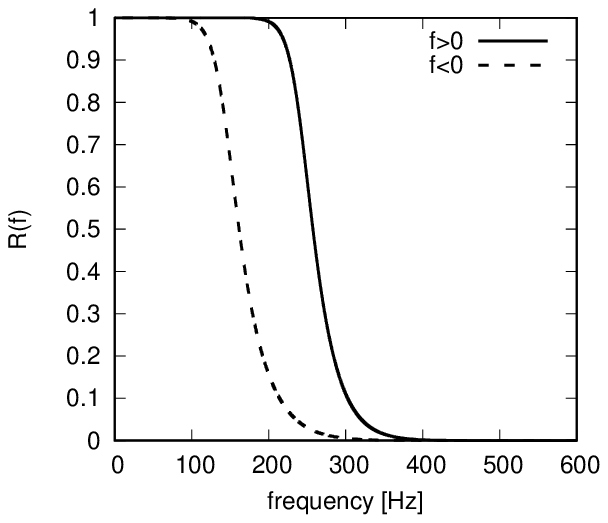}
\includegraphics[scale=0.95]{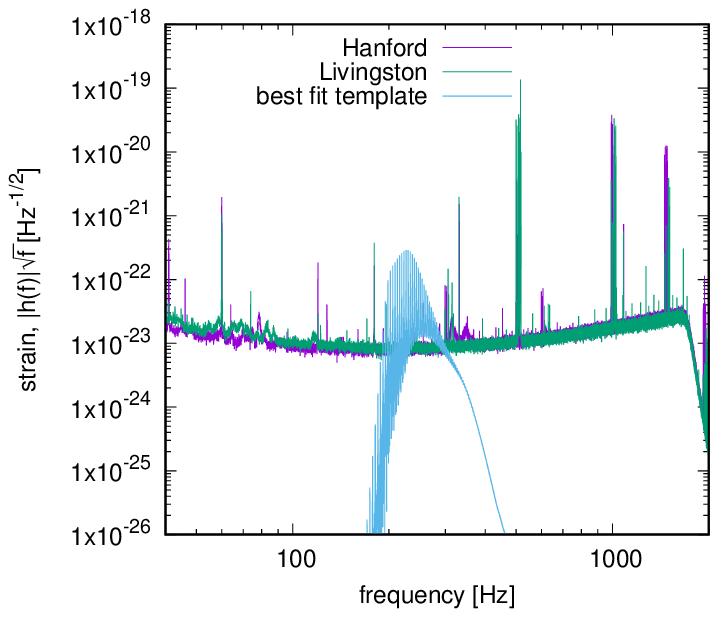}
\caption{Top panel: Reflection rate for the best fit values of $(a,M)$ for GW150914. For the $f<0$ case, the $x$ axis is adjusted to $f>0$. Bottom panel: Spectrum of the best fit echo template for GW150914 and the estimated amplitude spectral densities for Hanford and Livingston detectors. }
\label{template_spec}
\end{figure}

As an example, we show the reflection rate and the spectrum of the best fit template in our analysis and estimated average spectral densities for Hanford and Livingston detectors for GW150914 in Fig.~\ref{template_spec}.
From the bottom panel of Fig.~\ref{template_spec}, we can see that the amplitude of the template at a lower frequency is suppressed compared to the template given in \cite{abedi}.
This is because the transmission rate at the barrier is included in the template waveform of Eq.~\eqref{template1}.
Since $R(f)\sim1$ for $ f < 200\,\rm Hz$ in the case of Fig.~\ref{template_spec}, echoes at those frequencies are strongly suppressed.

We further assume  an unknown dissipative effect so that the superradiant amplification does not occur, which is too small to affect our analysis, though.  

\section{Method of analysis}
\subsection{Analyses to search for echo signals}
We first search for the echo signals right after the binary black hole merger.
We use a matched filter analysis to evaluate the SNR $\rho$ defined as 
\begin{equation}
\rho = (x|h) =4 \mbox{Re} \left ( \int^{f_{\rm max}}_{f_{\rm min}} \frac{\tilde{x} (f) \tilde{h}^* (f)}{S_n(f)} df \right ),
\label{snr}
\end{equation}
where $\tilde{x} (f)$ is the Fourier transformation of the observed data, $\tilde{h} (f)$ is the template in the frequency domain, and $S_n(f)$ is the noise power spectrum of a detector.
 We set $f_{\rm max}=2048$ Hz and  $f_{\rm min}=40 $ Hz, and we normalize the template so that $(h|h) =1$. 
We use the first 1024 seconds of the 4096-second data for each event to estimate the noise power spectrum using Welch's method \cite{welch,allen}.
The KAGRA Algorithmic Library (\verb+KAGALI+) \cite{oohara} is used to estimate the noise power spectrum and to calculate the matched filter SNR.
Theoretically, the first echo should be at a specific time length from the merger.
As described in \cite{abedi}, we search for the maximum value of SNR in the range
\begin{equation}
0.99 \le T \equiv (t_{\rm echo}-t_{\rm merger})/\Delta t_{\rm echo} \le 1.01,
\label{interval}
\end{equation} 
where $ t_{\rm echo}$ is the starting time of the first echo.
The merger time of the binary black hole $t_{\rm merger}$ is determined by analyzing each event by the inspiral-merger-ringdown waveform $h_{ \mbox{{\tiny IMR}} }(t)$, and  $t_{\rm echo}$ such that the SNR becomes maximum in the time interval of Eq.~\eqref{interval} is determined by a matched filter analysis.
We search for the best fit values by varying three template parameters $(a,M,\phi_0)$.
The search regions of $(a,M)$ are 90\% credible regions estimated by the LIGO and Virgo collaborations \cite{ligo,gw170104} for GW150914, GW15012, GW151226, and GW170104, and by our reanalysis using the \verb+LALInference+ module within the LIGO Algorithmic Library (LAL) for the other events. 
Those explicit regions are shown in Table \ref{aM}.
Similarly, to obtain the inspiral-merger-ringdown template $h_{ \mbox{{\tiny IMR}} }(t)$ used in Eq.~\eqref{seed}, we use the values given by the LIGO tutorial \cite{tutorial} for the above four events and the values that give the maximum of the posterior probability given from our reanalysis using \verb+LALInference+ for the other events \footnote{Strictly speaking, we should change $h_{ \mbox{{\tiny IMR}} }(t)$ when we vary $(a,M)$.}.
The signal-to-noise ratio is also maximized for the initial phase of the template $\theta_{\rm ini}$, which can be obtained automatically by orthogonal templates for each $(a,M,\phi_0)$.
This is not considered in the previous studies \cite{abedi,westerweck}.

We use data from the Hanford and Livingston detectors.
To evaluate the network SNR, we sum the square of SNRs of respective detectors. 
This means that we basically perform a single detector search.
\begin{table}[h]
\begin{center}
\begin{tabular}{l  c c }
\hline
Event & $a/M$ &  $M/ M_{\odot}$  \\  \hline
GW150914    &$0.61-0.73$   &$64.2-71.8 $  \\
GW151012   &$0.55-0.74$   & $39.0-60.0$ \\
GW151226  &$0.67-0.78$   & $21.1-30.7 $  \\
GW170104  &$0.56-0.74$    &$52.3-63.2$  \\
GW170608 &$0.51-0.74$ &$52.0-61.4$ \\
GW170729&$0.73-0.89$  & $107.7-138.0$\\
GW170814   & $0.65-0.75$   &$55.8-61.5$     \\ 
GW170818  & $0.55-0.75$ &$64.7-79.1 $ \\
GW170823  &$0.57-0.79$ & $74.0-97.1 $\\ 
 \hline
 \end{tabular}
\caption{The search regions for $(a,M)$ in the detector frame, which correspond to 90\% credible regions by parameter estimations from binary black hole events.}
\label{aM}
\end{center}
\end{table}

\subsection{Background estimation and data}
Background estimation is necessary to evaluate the significance of the candidate obtained at the event data segment.
We follow the method given in Ref.~\cite{westerweck}.
We divide the 4096-second data into 32-second data segments and perform the same analysis shown in the previous subsection for all remaining data segments.
Then, we count the number of data segments which give the same or higher SNR obtained in the event data segment, 
and the p-value is defined as the ratio to the number of all segments.

There are two versions of LIGO open data for noise subtraction data, C01 and C02 \cite{data}.
Since the data for all observed events in O1 and O2 are given in the C02 version, it may be reasonable to use only C02 data.
However, 4096-second data of the Hanford detector is not available for GW151226.
For comparison with previous works, we also use C01 for four events,  GW150914, GW151012, GW151226, and GW170104.
Note that for GW170809, 4096-second data are not  available for the Livingston detector, so we do not include this event.
That is, we analyze nine binary black hole merger events observed in O1 and O2.

As mentioned in Ref.~\cite{westerweck}, to use 4096-second data for the background estimation, the data quality should be homogeneous throughout the period.
It is confirmed that the variations of data quality are small for GW150914, GW151012, GW151226, and GW170104 in Ref.~\cite{westerweck}, and we confirm small variations of the noise level for the other five events.

Also, as mentioned in Ref.~\cite{westerweck}, a short transient noise feature is observed in the beginning of the data of GW151012.
Therefore, we exclude some data located in the beginning of 4096-second data.
The total number of the reference data segments is 127 for all events.

We use a Tukey window with a parameter $\alpha=1/8$ to cut off the edges of time series data for all segments.
Since we do not want to lose the expected echo signals by the window function, we set the merger time at around 8 seconds from the beginning of a 32-second data segment for the event segments.

\section{Results}
We summarize the results of p-values in Table \ref{pvalue1}.
The results are divided into two data versions, C01 and C02.
A hyphen means that 4096-second data are not available.
We set the critical p-value as 0.05, which corresponds to roughly 2$\sigma$ significance.
In our case, if the p-value is below (above) the value, then echo signals are likely (unlikely) to be present in the data.
Our results show that p-values for all events and the combined p-value well exceed this critical value; that is, echo signals modeled within our framework do not exist in the data, or the amplitude of the signals are too small to be detected within the current detector sensitivity.
We also confirm that the variation of $t_0$ weakly affects SNR; therefore, fixing $t_0 = -0.1 \Delta t_{\rm echo}$ is a reasonable assumption to save computational costs. 

In our analysis, we also consider the best fit of the initial phase of the template $\theta_{\rm ini}$, which is different from the previous studies \cite{abedi, westerweck}, so it might be inappropriate to compare the results directly.
However, we also analyze echo signals using the same template as in Abedi et al.~\cite{abedi} and probably with the same condition for the analysis, the results and comparison to those given by Westerweck et al.~\cite{westerweck} are shown in Appendix~\ref{follow}.
We additionally analyze the O2 events with this template, which gives similar p-value as that of O1 events. 
Results are shown in Appendix~\ref{O2}. 

\begin{figure}[h]
\includegraphics[scale=1.1]{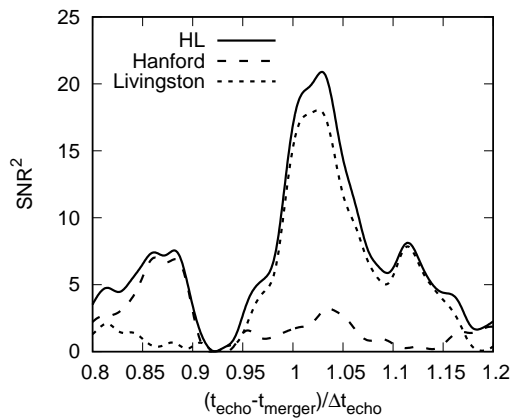}
\caption{Square of signal-to-noise ratio against $T \equiv (t_{\rm echo}-t_{\rm merger})/\Delta t_{\rm echo} $.
Solid, dashed, and dotted lines correspond to $\rho^2$ for combined (Hanford and Livingston), Hanford, and Livingston, respectively, for the best fit parameter case for GW150914.  }
\label{rho_bestfit}
\end{figure}

We show the detail of the behavior of SNR in Fig.~\ref{rho_bestfit} for the case of the best fit parameters of GW150914 (C01) as an example.
Solid, dashed, and dotted lines correspond to $\rho^2$ for combined (Hanford and Livingston), Hanford, and Livingston, respectively.
We can see a peak for the combined and Livingston cases near $T \sim 1$; however, the peak of the Hanford case is located slightly outside the interval of Eq.~\eqref{interval}.
The figure shows that $\rho^2$ oscillates slowly against $T$ compared to Fig.~7 in Ref.~\cite{abedi} because we consider the best fit initial phase of the template as well.

To see the effect of including a frequency-independent phase shift for the reflections as a parameter, we also analyze the case when only the phase inversion is considered for C01 data.
The results are given in Appendix~\ref{phase}.
The significance becomes lower if the phase shift is not fixed, except for GW151226.

\begin{table}[h]
\begin{center}
\begin{tabular}{l l l}
\hline
& \multicolumn{2}{c}{Data version} \\
Event & C01 & C02 \\  \hline
GW150914   & 0.992  & 0.984 \\
GW151012  & 0.646   &  0.882 \\
GW151226  & 0.276  &  -  \\
GW170104  & 0.717   &  0.677  \\
GW170608 &- &0.488 \\
GW170729&- & 0.575\\
GW170814  & - & 0.472     \\ 
GW170818 & -&0.976 \\
GW170823 &- & 0.315\\ \hline
Total & 0.976 &  0.921\\
 \hline
 \end{tabular}
\caption{P-values for each event and total p-value. A hyphen means that 4096-second of data are not available.}
\label{pvalue1}
\end{center}
\end{table}

\section{Conclusions and discussions}
We have searched for gravitational wave echo signals for nine binary black hole merger events observed by advanced LIGO and Virgo during the first and second observation runs.
We assume that the spacetime is entirely Kerr spacetime except that a reflective membrane is located near the event horizon radius.
We use the template waveform given by Nakano et al.~\cite{nakano}, in which the reflection rate and the phase shift at the potential barrier due to the angular momentum are calculated from Teukolsky equations. 
We assume a perfect reflection at the membrane; however, the phase shift at the membrane due to reflection is model dependent, so we assume the frequency-independent phase shift at both the membrane and the potential barrier as a parameter.
The transmission rate given from the reflection rate strongly suppresses the lower frequencies contained in the template waveform.
In addition to the echo parameters, we maximized the signal-to-noise ratio against the initial phase of the template.
We used  adjacent 4096-second data from open LIGO data for the background estimation, and evaluated the significance by p-values.
We found no significant echo signals within our analysis.
Since the method of analysis is slightly different from the analyses in the previous studies \cite{abedi,westerweck}, we cannot compare our results to theirs directly, but our results suggest that the suppression of the lower frequency part in the template may affect the p-value.

As mentioned in Introduction, previous injection studies \cite{westerweck, nielsen, lo} show that if the amplitude of the first echo is larger than 15\% of the peak amplitude of the binary black hole merger, echo signals can be detected by the current detector sensitivity, assuming a frequency-independent reflection rate.
Combined with their studies, our results suggest that the amplitude of echo signals should be much smaller than the peak amplitude of the merger even if echo signals exist.

However, p-values are much smaller when we use the template given by Abedi et al.~for both O1 events and O2 events.
This may imply that the waveform of their template is more favored than that of the template in our analysis,  
although our assumption is physically appropriate if we assume Kerr spacetime. 
Signals similar to the template of Abedi et al.~might be produced from non Kerr spacetime or unknown exotic physics, or instrumental reactions of the detector.  

The third LIGO and Virgo run started in April 2019, and about ten candidates for binary black hole mergers have been observed so far in the first two months \cite{gcn}.
Some of them might have higher SNR than the events observed in O1 and O2, which may enable us 
to detect echoes or to constrain their amplitude further.
To do so, it may be useful to coherently analyze the data from more than two detectors, besides improving the echo template.

\section*{Acknowledgments}
This research has made use of data, software, and web tools 
obtained from the Gravitational Wave Open Science
Center (https://www.gw-openscience.org), a service
of LIGO Laboratory, the LIGO Scientific Collaboration,
and the Virgo Collaboration. LIGO is funded by the
U.~S.~National Science Foundation. Virgo is funded by
the French Centre National de la Recherche Scientique
(CNRS), the Italian Istituto Nazionale di Fisica Nucleare
(INFN), and the Dutch Nikhef, with contributions by
Polish and Hungarian institutes.
This work was supported by JSPS KAKENHI Grant No.~JP17H06358 and also Grant No.~JP17H06357.
H.~N.~acknowledges support from JSPS KAKENHI Grant No.~JP16K05347.
T.~N.~is
supported in part by a Grant-in-Aid for JSPS Research
Fellows.
N.~S.~acknowledges support from JSPS KAKENHI Grant No.~JP16K05356.
T.~T.~acknowledges support from JSPS KAKENHI Grant No.~JP15H02087.



\begin{thebibliography}{9}
\bibitem{gw150914}
B. P. Abbot et al. (LIGO Scientific and  Virgo Collaborations), Phys. Rev. Lett. {\bf116}, 061102 (2016).

\bibitem{gw151226}
B. P. Abbot et al. (LIGO Scientific and  Virgo Collaborations), Phys. Rev. Lett. {\bf116}, 241103 (2016).

\bibitem{gw170104}
B. P. Abbot et al. (LIGO Scientific and  Virgo Collaborations), Phys. Rev. Lett. {\bf118}, 221101 (2017).

\bibitem{gw170814}
B. P. Abbot et al. (LIGO Scientific and  Virgo Collaborations), Phys. Rev. Lett. {\bf119}, 141101 (2017).

\bibitem{catalog}
LIGO Scientific and  Virgo Collaborations, arXiv:1811.12907.

\bibitem{echeverria}
F. Echeverria, Phys. Rev. D {\bf 40}, 3194 (1989).

\bibitem{dreyer} O. Dreyer, B. Kelly, B. Krishnan, L Samuel Fin and R. Lopez-Aleman, Classical Quantum Gravity \textbf{21}, 787 (2004).


\bibitem{ligo3} 
  B.~P.~Abbott {\it et al.} (LIGO Scientific and Virgo Collaborations),
  Phys.\ Rev.\ Lett.\  {\bf 116}, 221101 (2016); {\bf 121}, 129902(E) (2018).

\bibitem{giesler}
M. Giesler, M. Isi, M. Scheel, and S. Teukolsky, arXiv:1903.08284.

\bibitem{isi}
M. Isi, M. Giesler,  W. M. Farr, M. Scheel, and S. Teukolsky, arXiv:1905.00869.


\bibitem{cardoso}
V. Cardoso, E. Franzin, and P. Pani, Phys. Rev. Lett {\bf 116 }, 171101 (2016); {\bf 117} 089902(E) (2016).

\bibitem{cardoso2}
V. Cardoso, S. Hopper, C. F. B. Macedo, C. Palenzuela, and P. Pani,  Phys. Rev. D {\bf 94 }, 084031 (2016).


\bibitem{mazur}
P. O. Mazur and E. Mottola, Proc. Natl. Acad. Sci. U.S.A. \textbf{101}, 9545 (2004).

\bibitem{almheiri}
A. Almheiri, N. Marolf,  J. Polchinski,  and J. Sully, J. High. Energy Phys.  02, (2013) 062.


\bibitem{cardoso4}
V. Cardoso and P. Pani, Living Rev.~Relativity {\bf 22}, 4 (2019).


\bibitem{abedi}
J. Abedi, H. Dykaar, and N. Afshordi, Phys. Rev. D {\bf 96}, 082004 (2017).

\bibitem{ashton}
G. Ashton, O. Birnholtz, M. Cabero, C. Capano, T. Dent, B. Krishman, G. D. Meadors, A. B. Nielsen, A. Nitz, and J. Westerweck, arXiv:1612.05625.

\bibitem{westerweck}
J. Westerweck, A. B. Nielsen, O. Fischer-Birnholtz, M. Cabero, C. Capano, T. Dent, B. Krishnan, G. Meadors, and A. H. Nitz, Phys. Rev. D {\bf 97}, 124037 (2018).


\bibitem{nielsen}
A. B. Nielsen, C. D. Capano, O. Birnholtz and J. Westerweck, Phys. Rev. D {\bf 99} 104012 (2019).

\bibitem{lo}
R. K. L. Lo, T. G. F. Li, and A. J. Weinstein, Phys. Rev. D {\bf 99} 084052 (2019).
\bibitem{tsang}
K.~W.~Tsang, M. Rollier, A. Ghosh, A. Samajdar, M. Agathos, K. Chatziioannou, V. Cardoso, G. Khanna, and C. Van Den Broeck,
  Phys.\ Rev.\ D {\bf 98},  024023 (2018).

\bibitem{tsang2} 
  K.~W.~Tsang, A.~Ghosh, A.~Samajdar, K.~Chatziioannou, S.~Mastrogiovanni, M.~Agathos and C.~Van Den Broeck,
  arXiv:1906.11168.

\bibitem{nakano}
H. Nakano, T. Tanaka, N. Sago, and H. Tagoshi, Prog. Theor. Exp. Phys. {\bf 2017}, 071E01 (2017).

\bibitem{mark}
Z. Mark, A. Zimmerman,  S. M. Du, and Y. Chen, Phys. Rev. D {\bf 96}, 084002 (2017).

\bibitem{testa}
A. Testa and P. Pani, Phys. Rev. D \textbf{ 98}, 044018 (2018).

\bibitem{maggio} 
  E.~Maggio, A.~Testa, S.~Bhagwat and P.~Pani,
  arXiv:1907.0309.




\bibitem{maselli}
A. Maselli, S. H. V\"olkel, and K. D. Kokkotas, Phys. Rev. D {\bf 96}, 064045 (2017). 

\bibitem{ytwang}
Y-T. Wang, Z-P. Li, J. Zhang, S-Y. Zhou, and Y-S. Piao, Eur. Phys. J. C {\bf 78} (2018).


\bibitem{ytwang2}
Y-T. Wang, J. Zhang, S-Y. Zhou, and Y-S. Piao, arXiv:1904.00212.


\bibitem{cardoso3}
V. Cardoso, V. F. Foit, and M. Kleban, J.~Cosmol.~Astropart.~Phys.~08 (2019) 006.

\bibitem{wang}
Q. Wang, N. Oshita, and N. Afshordi, arXiv:1905.00446.

\bibitem{oshita}
N. Oshita, Q. Wang, and N. Afshordi, arXiv:1905.00464.

\bibitem{welch}
P. D. Welch, IEEE Trans. Audio Electroacoust. \textbf{ 15}, 70 (1967). 
\bibitem{allen}
B. Allen, W. G. Anderson, P. R. Brady, D. A. Brown, and J. D. E. Creighton, Phys. Rev. D \textbf{85}, 122006 (2012).

\bibitem{oohara}
K. Oohara {\it et al.} in {\it Proceedings of the14th Marcel Grossmann Meeting on Recent
                        Developments in Theoretical and Experimental General
                        Relativity, Astrophysics, and Relativistic Field Theories
                        (MG14) (In 4 Volumes): Rome, Italy, 2015}, (World Scientific, 2017), Vol.~3, p.~3170.


\bibitem{ligo}
B. P. Abbot et al. (LIGO Scientific and  Virgo Collaborations), Phys. Rev. X {\bf 6}, 041015 (2016).

\bibitem{tutorial}
\verb|https://www.gw-openscience.org/tutorials/.|

\bibitem{data}
\verb|https://www.gw-openscience.org.|


\bibitem{gcn}
\verb|https://gcn.gsfc.nasa.gov/gcn3_archive.html.|



\end{thebibliography}
%

\appendix

\vspace{ 3pt}
\section{Analysis using the template of Abedi et al.}
In this appendix, we show the results using the same template given in Abedi et al.~\cite{abedi}.
Here we fix the cutoff parameter $t_0$ as described in Sec.~II, and we set $\Delta t_{\rm echo}$ and a frequency-independent reflection rate $\gamma$ as free parameters.
The initial phase of the template is fixed to zero.
\subsection{O1 events (reanalysis of Westerweck et al.) }
\label{follow}
Since we follow Westerweck et al.~\cite{westerweck} for the background estimation, it would be appropriate to compare our results with theirs.
Table \ref{pvalue3} shows the results of p-values for three O1 events.
The results of Westerweck et al.~are denoted as AEI.
The Poisson errors of the p-values for GW151226 and GW151012 are not given in \cite{westerweck}, so we estimate the errors from the p-value and the number of segments they use.
We can see that both results are almost consistent within the Poisson errors for all events.
Since we use a 32-second template while Westerweck et al.~only show the results of a 16-second template for GW170104,  we do not compare the results of this event here.

\begin{table}[h]
\begin{center}
\begin{tabular}{l l l}
\hline
Event &AEI \cite{westerweck}& Ours \\  \hline
GW150914   & 0.238 $\pm$ 0.043  & 0.157 $\pm$ 0.035\\
GW151012  & 0.063 $\pm$ 0.022   &  0.047 $\pm$ 0.019 \\
GW151226  & 0.476 $\pm$ 0.061  &  0.598 $\pm$ 0.069 \\ \hline
Total  & 0.032 $\pm$ 0.016  &  0.055 $\pm$ 0.021 \\
 \hline
 \end{tabular}
\caption{P-values and Poisson errors for O1 events.}
\label{pvalue3}
\end{center}
\end{table}

\subsection{O2 events}
\label{O2}
We also the data analyze for O2 events.
For GW170104, we use C01 data and for the other events, we use C02 data.
We set the search region of $\Delta t_{\rm echo}$ from the 90\% credible regions of $(a,M)$ as described in Sec.~III A.
P-values are given in Table \ref{pvalue4}.
As shown in Table \ref{pvalue4}, the total p-value for the six O2 events is 0.039, while when combining with O1 events shown in Table \ref{pvalue3}, the p-value for nine events becomes 0.047.
\begin{table}[h]
\begin{center}
\begin{tabular}{l l}
\hline
Event & \\  \hline
GW170104   & 0.071  \\
GW170608  & 0.079    \\
GW170729  & 0.567  \\
GW170814  & 0.024    \\
GW170818  & 0.929   \\
GW170823  & 0.055 \\ \hline
Total  & 0.039   \\
 \hline
 \end{tabular}
\caption{P-values for O2 events.}
\label{pvalue4}
\end{center}
\end{table}

\section{Effect of the phase shift due to the reflection}
\label{phase}
As mentioned in Sec. II, the phase shift at the potential barrier can be calculated numerically.
However, since the phase shift at the membrane is model dependent, it is physically reasonable to assume the total phase shift as a parameter.
In the previous studies \cite{abedi, westerweck}, only phase inversion at the membrane is considered.
So in this section, we compare the results of two cases, when the phase shift is fixed to $\pi$ (result 1) and when it is a free parameter (result 2), respectively, in Table \ref{pvalue2}.
The template given in Eq.~\eqref{template1} is used.
Except for GW151226, p-values become slightly larger when the phase shift due to the reflection is left as a free parameter, which we believe is a more physical condition.

\begin{table}[h]
\begin{center}
\begin{tabular}{l l l}
\hline
Event & Result 1 & Result 2 \\  \hline
GW150914   & 0.638  & 0.992 \\
GW151012  & 0.417   &  0.646  \\
GW151226  & 0.953  &  0.276  \\
GW170104  & 0.213   &  0.717  \\  \hline
Total & 0.528 &  0.976 \\
 \hline
 \end{tabular}
\caption{P-value for each event and total p-value. 
Result 1 is the case when the phase shift is fixed to $\pi$, and result 2 is the case when the total phase shift is also a parameter. }
\label{pvalue2}
\end{center}
\end{table}

\end{document}